\documentclass[preprint,12pt]{elsarticle}
\usepackage{graphicx}
\usepackage{epsfig}
\usepackage{amssymb,amsmath}
\usepackage{multicol}
\RequirePackage[dvipsnames,usenames]{color}
\journal{Physics Letters B}
\newcommand{\mpl}{m_{\rm P}}
\begin{document}
\begin{frontmatter}
\title{\textbf{Graviton propagator, renormalization scale and black-hole like states}}
\author{X.~Calmet$^{1}$}\ead{x.calmet@sussex.ac.uk}
\author{R.~Casadio$^{2}$}\ead{Roberto.Casadio@bo.infn.it}
\author{A.Yu.~Kamenshchik$^{2,3}$}\ead{Alexander.Kamenshchik@bo.infn.it}
\author{O.V.~Teryaev$^{4,5}$}\ead{teryaev@theor.jinr.ru}  
\address{\small $^1$Department of Physics $\&$ Astronomy, University of Sussex,
\\
Falmer, Brighton, BN1 9QH, United Kingdom. 
\\
\small $^2$Dipartimento di Fisica e Astronomia, Universit\`a di Bologna, and INFN,
\\
Via Irnerio 46, 40126 Bologna, Italy.
\\
\small $^3$L.D. Landau Institute for Theoretical Physics of the Russian Academy of Sciences,
\\
Kosygin str. 2, 119334 Moscow, Russia.
\\
\small $^4$Bogoliubov Laboratory of Theoretical Physics, Joint Institute for Nuclear Research, 
\\
141980 Dubna, Russia.
\\
\small $^5$ Lomonosov Moscow State University,
\\
Leninskie Gory 1, 119991, Moscow, Russia}
%
\date{ \ }
\begin{abstract}
We study the analytic structure of the resummed graviton propagator, inspired by the possible
existence of black hole precursors in its spectrum.
We find an infinite number of poles with positive mass, but both positive and negative effective
width, and studied their asymptotic behaviour in the infinite sheet Riemann surface. 
We find that the stability of these precursors depend crucially on the value of the normalisation
point scale.
\end{abstract}
\end{frontmatter}
\section{Introduction}
Propagators play a crucial role in both quantum mechanics (see e.g.~\cite{Holstein})
and in quantum field theory (see e.g.~\cite{Bog-Shir,Ber-Lif-Pit}). 
As is well-known, the appearance of a pole in the free field propagator ($p^2=m^2$)
tells us that there exists a one-particle state with the corresponding mass $m$.   
The vacuum polarization loops summation for the photon propagator in quantum electrodynamics
revealed the existence of a particular pole at a huge negative value of $p^2$~\cite{AKL},
which is called the ``Landau pole''.
Because of gauge invariance, the existence of this pole implies the same pole in
the effective charge of the electron.
The latter can be removed by imposing causality and using some adequate analytic
properties of propagators~\cite{Redmond,Shirkov}. 
The generalisation of this procedure was later applied to quantum
chromodynamics (QCD)~\cite{Shirkov1}, resulting in the successful description of various
physical processes~\cite{anal-other}.
\par   
More recently, the resummed one-loop propagator of the graviton interacting with
matter fields was obtained~\cite{grav-prop,grav-prop1} (see also~\ref{A1}).
This propagator has a rather elegant, but involved  form, namely
\begin{equation}
i\,D^{\alpha\beta}(p^2)
=
i
\left(L^{\alpha\mu}L^{\beta\nu}+L^{\alpha\nu}L^{\beta\mu}-L^{\alpha\beta}L^{\mu\nu}\right)
G(p^2)
\ ,
\label{prop}
\end{equation}
where 
\begin{equation}
L^{\mu\nu}(p)
=
\eta^{\mu\nu}-\frac{p^{\mu}p^{\nu}}{p^2}
\label{prop1}
\end{equation}
and 
\begin{equation}
G^{-1}(p^2)
=
2\,p^2
\left[1-\frac{N\,p^2}{120\,\pi\, \mpl^2}\ln\!\left(-\frac{p^2}{\mu^2}\right)\right]
\ .
\label{prop2}
\end{equation}
Here, $\mpl$ denotes the Planck mass, $\mu$ is the renormalization scale,
$N=N_s+3\,N_f+12\,N_V$, where $N_s$, $N_f$, $N_V$
are the number of scalar, fermion and vector fields, respectively.
In the Standard Model, $N_s=4$, $N_f = 45$, $N_V=12$ and $N=283$.
The propagator~(\ref{prop}) has a standard pole at $p^2=0$ and an infinite
number of other poles, which are the zeros of the expression~(\ref{prop2}).
It was suggested that these poles correspond to the appearance of a sort of
precursors of quantum black holes in Refs.~\cite{Calmet,Casadio}.
\par
In this paper we shall study in detail the poles and discuss their possible physical interpretations.
In particular, we reveal a multi-sheet structure of the corresponding Riemann surface,
the role of the renormalisation point and some analogies with studies of the propagator
in QCD.
\section{Poles of the graviton propagator}
We shall now proceed to study the structure of the graviton propagator that follows
from the expression~\eqref{prop2}.
We shall in particular derive expressions for the mass and width, and analyse their
location in the Riemann surface.
\subsection{Pole positions}
It is convenient to rewrite the equation for the non-trivial zeros of~(\ref{prop2}) as 
\begin{equation}
z\ln z
=
- A
\ ,
\label{zeros}
\end{equation}
where the new variable $z$ is defined as 
\begin{equation}
z
\equiv 
-\frac{p^2}{\mu^2}
\label{z}
\end{equation}
and the positive constant  
\begin{equation}
A
=
\frac{120\, \pi\, \mpl^2}{N\, \mu^2}
\ .
\label{A}
\end{equation}
Note that~(\ref{zeros}) is nothing but the well known Lambert equation~\cite{Lambert}. 
Introducing as usual $z=\rho\, e^{i\,\theta}$,
we can rewrite~(\ref{zeros}) as a pair of equations for the imaginary and real
parts of the expression on the left-hand side, that is
\begin{equation}
\ln\rho\,\sin\theta+\theta\,\cos\theta
=
0
\ ,
\label{zeros1}
\end{equation}
\begin{equation}
\rho\,\ln\rho\,\cos\theta-\rho\,\theta\,\sin\theta
=
-A
\ .
\label{zeros2}
\end{equation}
First of all, let us consider the particular case when $\theta = 0$.
Then, Eq.~(\ref{zeros1}) is satisfied automatically, while Eq.~(\ref{zeros2}) takes the form
\begin{equation}
\rho\,\ln\rho
=
-A
\ .
\label{zeros20}
\end{equation}
This equation has two solutions if $A < 1/e$, which merge at $A = 1/e$. 
Since $z={\rm Re}( z) >0$, the real part of the corresponding pole for $p^2$ is negative and 
both these solutions are tachyons.
Obviously, they are stable because ${\rm Im}( p^2) = 0$. 
\par
\begin{figure}[t]
\centerline{\epsfxsize 8cm
\epsfbox{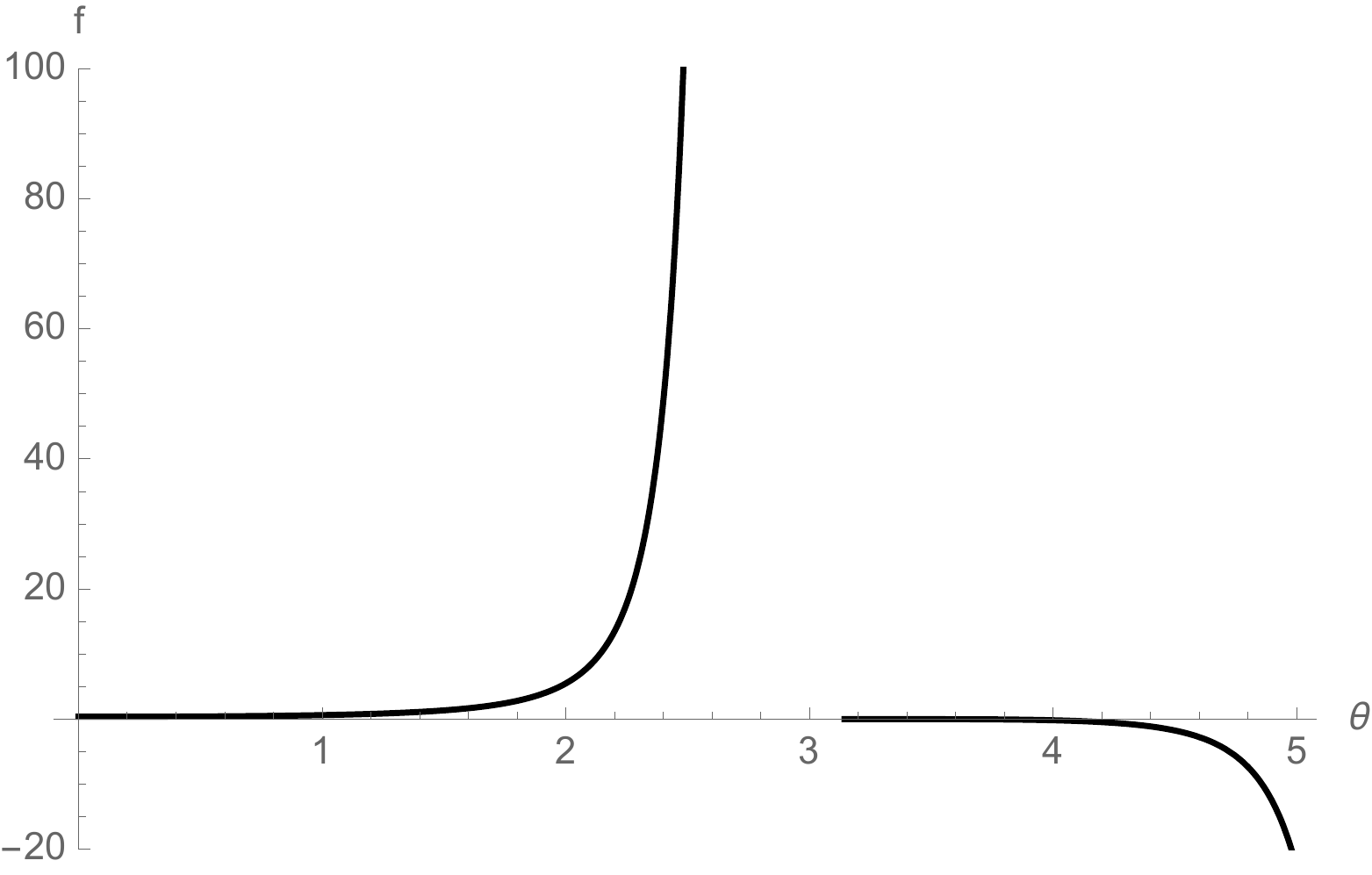}}
\caption{Function $f$ for $0<\theta<2\,\pi$.
Solutions of Eq.~\eqref{theta} can only exist in $0<\theta<\pi$ where $f>1/e>0$.
\label{ftheta}}
\end{figure}

 
If $\theta \neq n\,\pi$ (with $n$ integer), it follows from Eq.~(\ref{zeros1}) that
\begin{equation}
\ln \rho
=
-\frac{\theta}{\tan\theta}
\ .
\label{zeros3}
\end{equation}
Substituting~(\ref{zeros3}) into Eq.~(\ref{zeros2}) yields
\begin{equation}
\rho
=
\frac{A\sin\theta}{\theta}
\ .
\label{zeros4}
\end{equation}
Combining Eqs.~(\ref{zeros4}) and~(\ref{zeros3}), we obtain the equation for the phase $\theta$,
\begin{equation}
f(\theta)
\equiv
\frac{\theta}{\sin\theta}\,
\exp\left(-\frac{\theta}{\tan\theta}\right)
=
A
\ ,
\label{theta}
\end{equation}
where the function $f(\theta)$ is plotted for $0<\theta<2\,\pi$ in Fig.~\ref{ftheta}.
\begin{figure}[t]
\centerline{\epsfxsize 8cm
\epsfbox{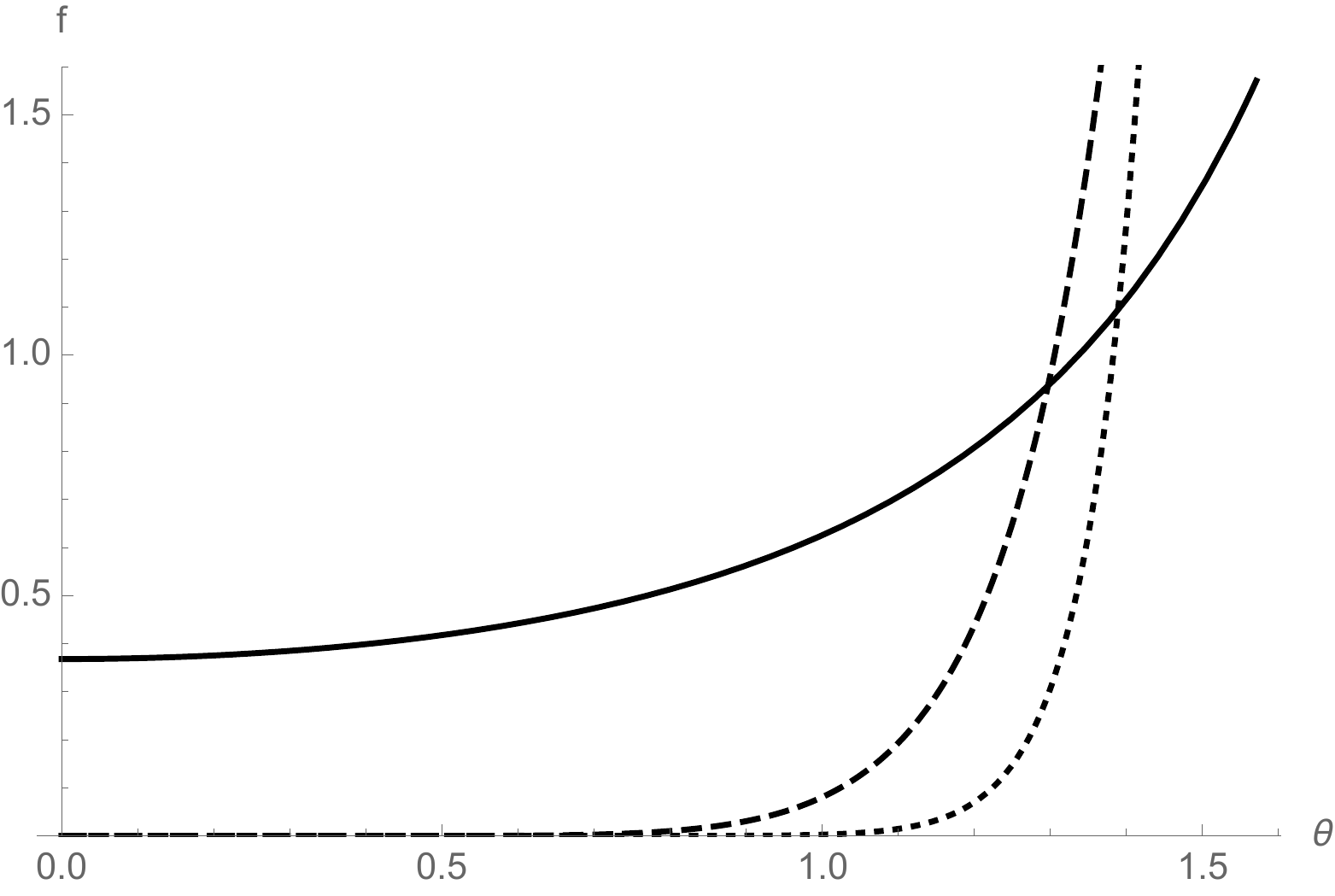}}
\caption{Function $f$ for $n=0$ with $0<\theta<\pi/2$ (solid line),
for $n=1$ with $2\,\pi<\theta<5\,\pi/2$ (dashed line) and
for $n=2$ with $4\,\pi<\theta<9\,\pi/2$
(dotted line).
Note the origin is shifted to $2\,n\,\pi$.
\label{f0f1}}
\end{figure}
\par
Obviously, we are interested only in solutions of Eq.~(\ref{theta}) which correspond to 
a positive $\rho$ from~(\ref{zeros4}).  
Note, however, that solutions of Eq.~(\ref{theta}) exist only if $\theta$ and $\sin\theta$ have
the same sign, in which case the right-hand side of Eq.~(\ref{zeros4}) is positive, as required.
Therefore, Eq.~(\ref{theta}) has solutions only in the intervals 
\begin{equation}
2\,\pi\, n
<
\theta
<
(2\,n+1)\,\pi
\ ,
\qquad
n
=
0,1,2,\ldots
\label{int}
\end{equation}
and in the mirror symmetric intervals with negative values of $\theta$.
At the same time, no solutions exist in the intervals 
\begin{equation}
(2\,n+1)\,\pi
\leq
\theta
\leq
(2\,n+2)\,\pi
\ ,
\qquad
n=0,1,2,\ldots
\label{noint}
\end{equation}
and in the mirror intervals. 
\par
Let us consider the behaviour of the function $f(\theta)$ in Eq.~(\ref{theta}) for $0\leq \theta< \pi$
(the interval~(\ref{int}) with $n=0$). 
In this interval, $f(\theta)$ grows monotonicallly from the minimum $f(0)={1}/{e}$ to the particular
value 
\begin{equation}
f\left(\frac{\pi}{2}\right) = \frac{\pi}{2}
\ ,
\label{function1}
\end{equation}
where $\rho(\pi/2) =1$,
and then diverges for $\theta \to \pi$.
That means that Eq.~(\ref{theta}) has no roots if $A \leq {1}/{e}$
(excluding the two real roots of Eq.~(\ref{zeros20}) already described above),
has one root in the interval $0 \leq \theta < {\pi}/{2}$ if ${1}/{e} \leq A < {\pi}/{2}$,
and one root in ${\pi}/{2} \leq \theta < \pi$ if $A\geq {\pi}/{2}$. 
It also implies that, for $A > {\pi}/{2}$, the equation for the phase has a unique root
with ${\pi}/{2} \leq \theta <\pi$, and the corresponding $z$ has a negative real part
and a positive imaginary part.
\par
As we are interested in the complete complex structure of the resummed graviton propagator,
we shall consider the whole Riemann surface with all of its sheets.
On the second sheet, the relevant interval is given by Eq.~(\ref{int}) with $n=1$, that is $2\,\pi < \theta < 3\,\pi$.
Here, the function~(\ref{theta}) grows from $0$ to ${5\,\pi}/{2}$ when the phase $\theta$ goes from $2\,\pi$ to
$5\,\pi/2$, and then keeps on growing indefinitely.
A comparison of $f(\theta)$ near its minimum at $2\,\pi\,n$ for $n=0,$ $1$ and $2$ is shown in Fig.~\ref{f0f1},
from which we see that the curve of $f(\theta)$ moves to the right and gets steeper for increasing $n$
(this trend continues for larger values of $n$).
It is worth remarking that this minimum is always $f(2\,\pi\,n)=0$ except for the fundamental sheet $n=0$,
where $f(0)=1/e$. 
In general, larger values of $A$ result in larger values of the phases $\theta_n$ of the pole positions $z_n$,
and the dependence of the phase $\theta_n$ on $n$ will be analysed in more detail below.
\subsection{Mass and width}
The poles of propagators in the complex domain correspond to unstable particles and this relation
is provided by the famous Breit-Wigner formula~\cite{BW}.
This formula may be presented in various ways \cite{wille,bohm}, all of which practically coincide
for non-relativistic  particles.
The covariant representation of the standard non-relativistic formula leads to the parametrization
for the pole position given by
\begin{eqnarray}
p^2
=\left(m-i\,\Gamma/2\right)^2
=
m^2-i\,\Gamma\,m-\Gamma^2/4
\ ,
\label{p2bohm}
\end{eqnarray}
which is suggested to be the preferable one (see Refs.~\cite{wille,bohm}). 
We will hence use this expression for the interpretation of the poles of the graviton propagator.
We would like to mention also the rather common expression 
\begin{eqnarray}
p^2
=
m^2-i\,\Gamma\,m
\ ,
\label{p2comm}
\end{eqnarray}
which is obviously close to Eq.~(\ref{p2bohm}) for narrow resonances with $\Gamma \ll  m$. 
\par
Note that the expression~(\ref{p2bohm}) is more suitable for investigating wide resonances.
In particular, a negative real part of $p^2$ can be reconciled with positive $m^2$, which is not the case for (\ref{p2comm}).
We shall dwell on this point in more detail.
Let us try to find the parameters $m$ and $\Gamma$ corresponding to the poles 
found in the previous subsection. 
Comparing the real and imaginary parts of Eq.~(\ref{p2bohm}) with those of the position of a pole,
we obtain
\begin{equation}
m^2-\frac{\Gamma^2}{4}
=
-\mu^2\,\rho_n\,\cos\theta_n
\ ,
\label{mGam}
\end{equation}
and
\begin{equation}
\Gamma
=\frac{\mu^2\,\rho_n\,\sin\theta_n}{m}
\ ,
\label{mGam2}
\end{equation}
where the index $n$ labels the sheet in Eq.~\eqref{int}.
On substituting Eq.~\eqref{mGam2} into Eq.~(\ref{mGam}), we obtain a quadratic equation for $m^2$.
The first solution is given by
\begin{eqnarray}
m_1^2
&\!\!=\!\!&
\mu^2\,\rho_n\,\sin^2\left(\frac{\theta_n}{2}\right)
\  ,
\nonumber
\\
\label{mGam3}
\\
\Gamma_1^2
&\!\!=\!\!&
4\, \mu^2\,\rho_n\,\cos^2\left(\frac{\theta_n}{2}\right)
\ ,
\nonumber
\end{eqnarray}
with $m_1^2$ and $\Gamma_1^2$ positive definite
and their ratio
\begin{equation}
\frac{\Gamma}{m}
=
2\, \cot\left(\frac{\theta_n}{2}\right)
\ .
\label{R}
\end{equation}
The second solution may also be obtained from the first one by interchanging $m \leftrightarrow 4\, i\,\Gamma$ 
and reads
\begin{eqnarray}
m_2^2
&\!\!=\!\!&
-\mu^2\,\rho_n\,\cos^2\left(\frac{\theta_n}{2}\right)
\  ,
\nonumber
\\
\Gamma_2^2
&\!\!=\!\!&
-4\, \mu^2\,\rho_n\,\sin^2\left(\frac{\theta_n}{2}\right)
\ ,
\label{mGam4}
\\
\frac{\Gamma_2}{m_2}
&\!\!=\!\!&
-2\, \tan\left(\frac{\theta_n}{2}\right)
\ ,
\nonumber
\end{eqnarray}
with $m_2^2$ and $\Gamma_2^2$ negative definite.
Obviously, the second solution~(\ref{mGam4}) implies an imaginary mass $m$ and an imaginary value of $\Gamma$.
Imaginary $m$ and $\Gamma$ make the appeal to the Breit-Wigner type of the representation for the poles in the
propagator meaningless, and we therefore completely discard Eq.~\eqref{mGam4} henceforth.
\par
Let us then consider the solution~\eqref{mGam3} and further require that $m$ is positive, 
so that  the corresponding expression for the width is given by
\begin{equation}
\Gamma
=
\frac{\mu\,\sqrt{\rho_n}\,\sin\theta_n}{\left|\sin({\theta_n}/{2})\right|}
\ .
\label{mGam5}
\end{equation} 
Note that to any positive value of $\theta_n$ there corresponds a mirror negative value.
Thus, we have pairs of poles with positive mass and both positive and negative values of the width $\Gamma$. 
In principle, it would also be possible to fix $\Gamma$ positive and similarly get a pair of positive and negative masses.
The appearance of such pairs of either $\Gamma$ or $m$ is simply related to the fact that Eq.~(\ref{theta}) is even in $\theta$.  
\par
Let us also remark that, had we chosen the representation~(\ref{p2comm}) for the poles of the propagator,
we would not be able to avoid the appearance of the negative mass squared and imaginary masses and widths for the poles
with $\cos\theta_n > 0$.
Thus, the choice of the representation (\ref{p2bohm}) and the solution~(\ref{mGam3}) looks justified in the case of
the graviton propagator. 
\begin{figure}[t]
\centerline{\epsfxsize 8cm 
\epsfbox{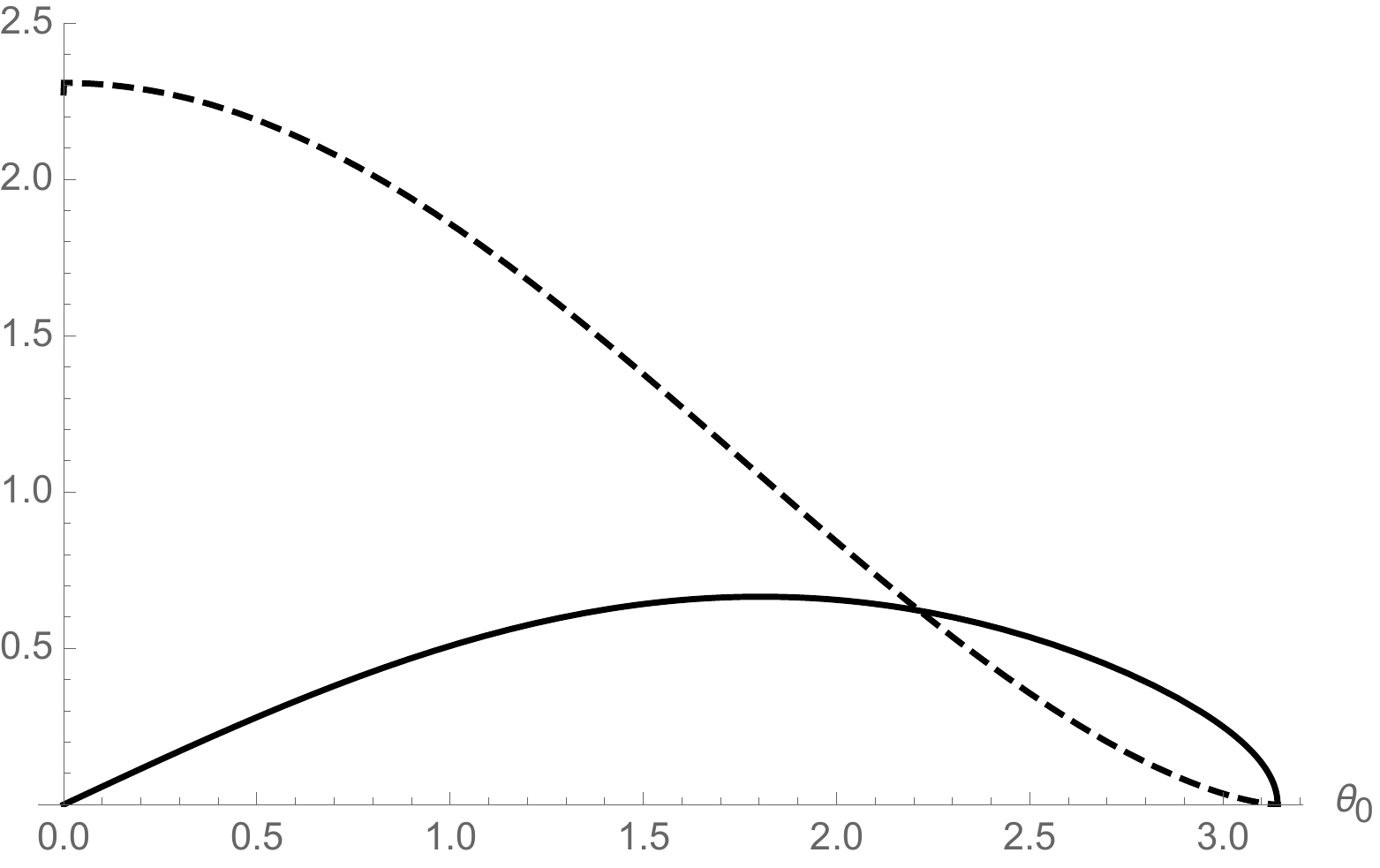}}
\caption{Mass $m$ (solid line) and width $\Gamma$ (dashed line) in units of $\mpl$ versus the pole phase
$0<\theta_0<\pi$.
\label{m0g0}}
\end{figure}
\begin{figure}[t]
\centerline{\epsfxsize 8cm 
\epsfbox{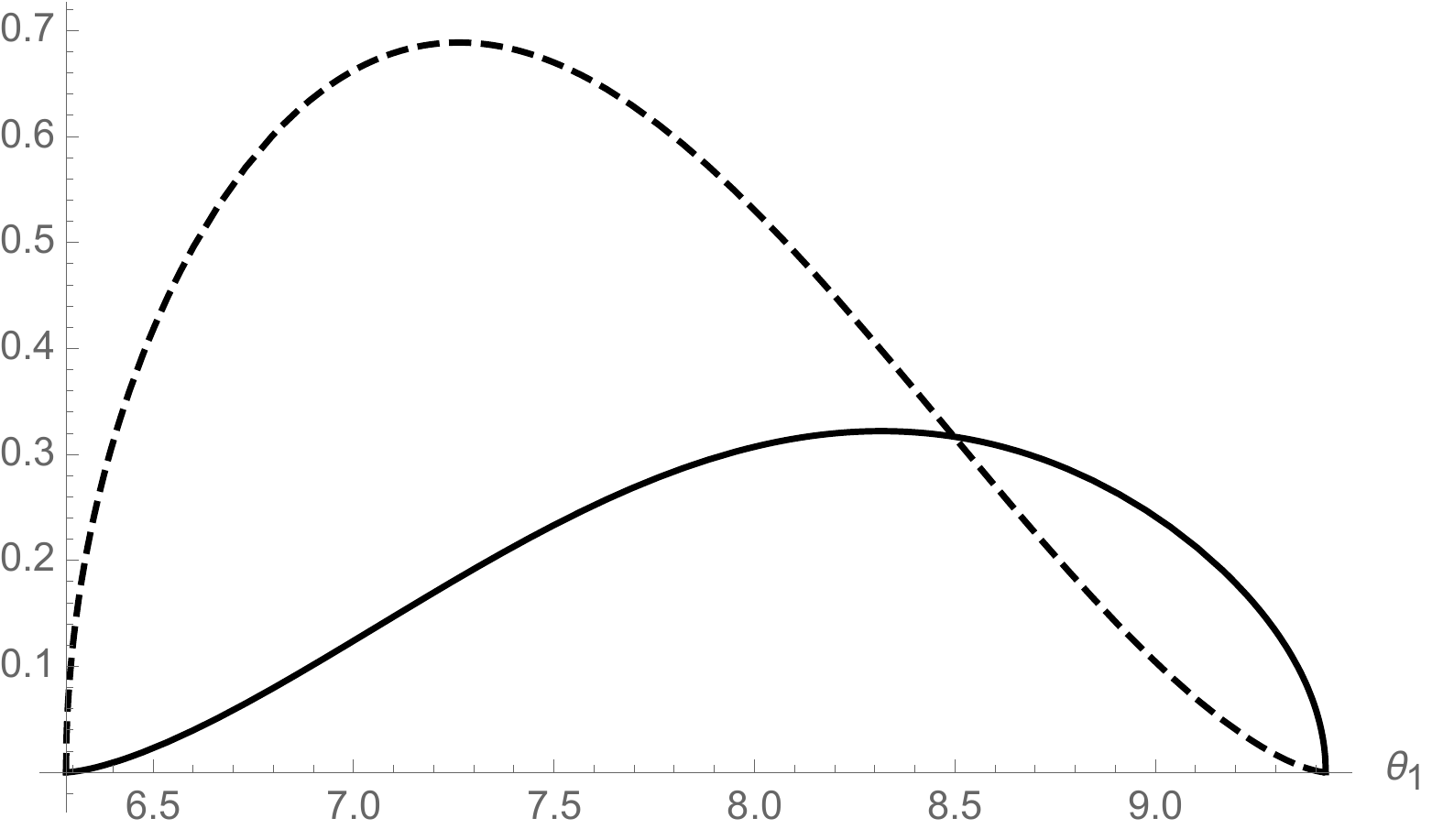}}
\caption{Mass $m$ (solid line) and width $\Gamma$ (dashed line) in units of $\mpl$ versus the pole phase
$2\,\pi<\theta_1<3\,\pi$.
\label{m1g1}}
\end{figure}
\par
Now, using Eqs.~(\ref{zeros4}) and (\ref{A}), we can rewrite the expression for the mass as
\begin{equation}
m
=
m_{\rm P}\,
\sqrt{\frac{120\,\pi}{N}\,\frac{\sin\theta_n}{\theta_n}}
\left|\sin\!\left(\frac{\theta_n}{2}\right)\right|
\ ,
\label{mGam6}
\end{equation}
and the width as
\begin{equation}
\Gamma
=
m_{\rm P}\,\sqrt{\frac{120\,\pi}{N}\,\frac{\sin\theta_n}{\theta_n}}\,
\frac{\sin\theta_n}{\left|\sin({\theta_n}/{2})\right|}
\ ,
\label{mGam7}
\end{equation}
and note that their ratio is again simply given by
\begin{equation}
\frac{\Gamma}{m}
=
2\,\cot\left(\frac{\theta_n}{2}\right)
\ ,
\label{GsM}
\end{equation}
Remarkably, these physical quantities depend only on the total number of fields $N$ and the phase
$\theta_n$ of the pole, but not on the renormalisation scale $\mu^2$.
We should however not forget that their very existence depends on the value of $A$ which
contains $\mu^2$.
\begin{figure}[t]
\centerline{\epsfxsize 8cm \epsfbox{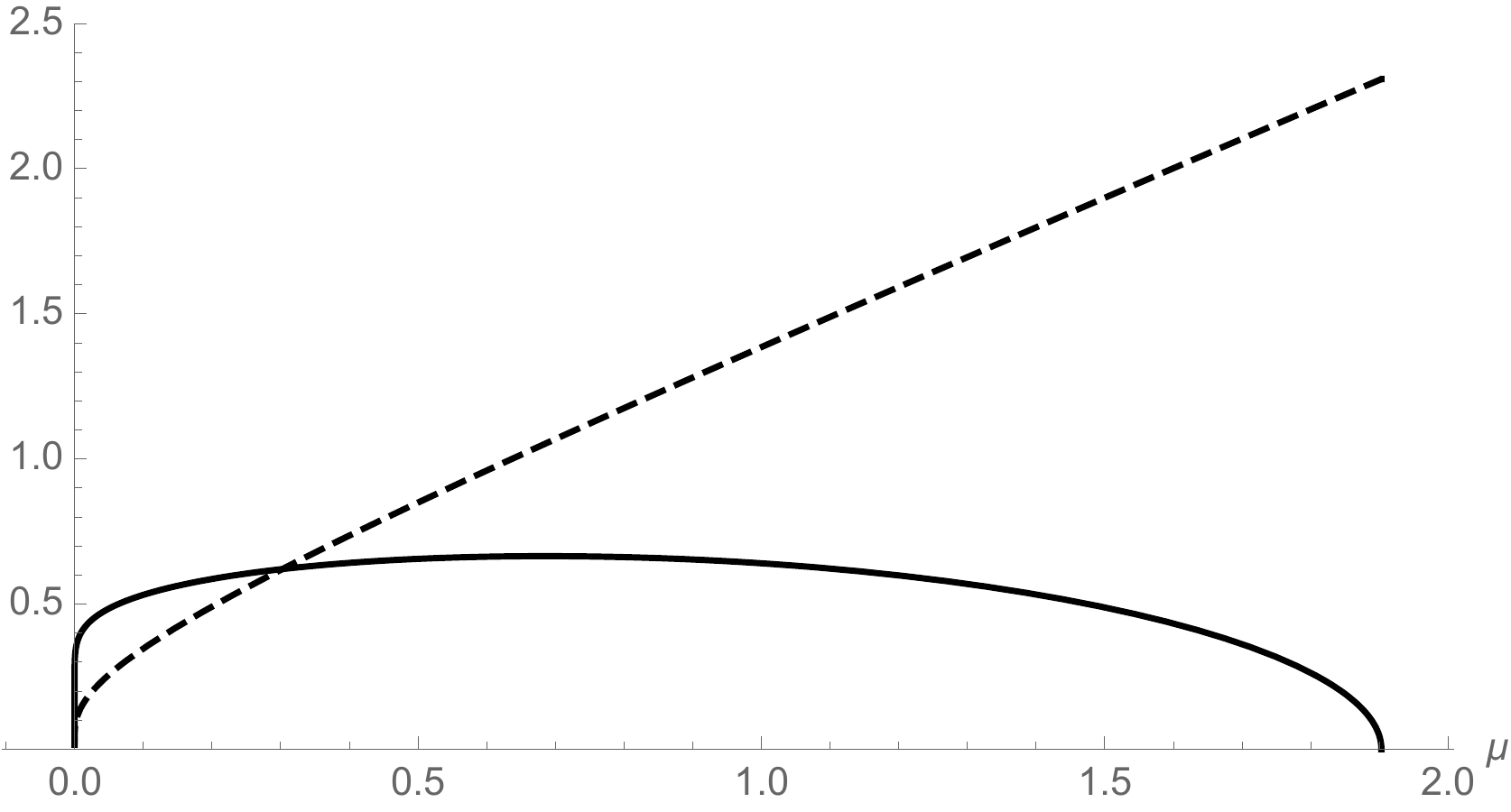}}
\caption{Mass $m$ (solid line) and width $\Gamma$ (dashed line) versus the renormalisation scale $\mu$
(in units of $\mpl$) for a pole in the fundamental sheet $n=0$. 
\label{mGamVmu0}}
\end{figure}
\begin{figure}[t]
\centerline{\epsfxsize 8cm \epsfbox{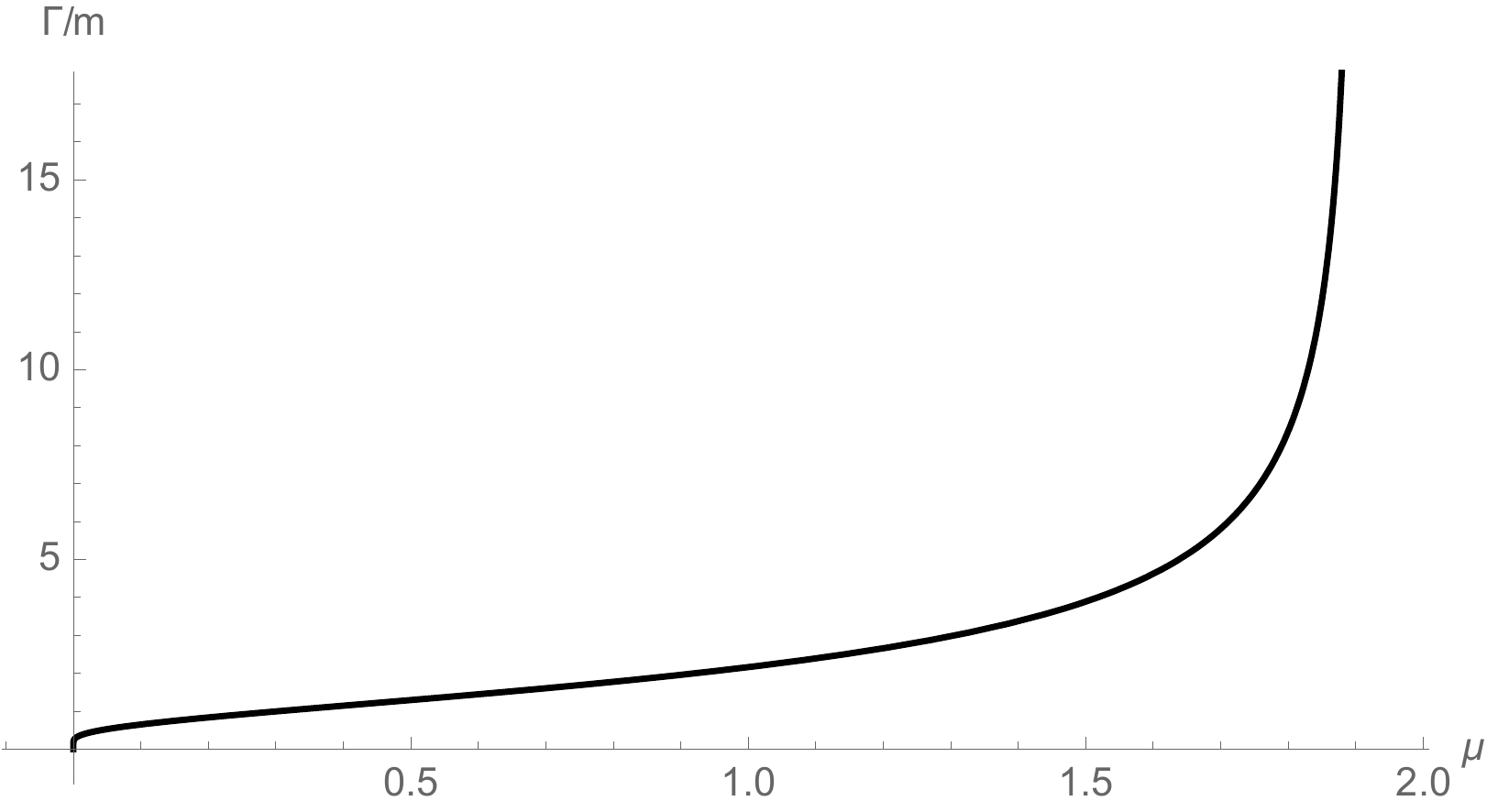}}
\caption{Ratio $\Gamma/m$ versus the renormalisation scale $\mu$
(in units of $\mpl$) for a pole in the fundamental sheet $n=0$. 
\label{RatioVmu0}}
\end{figure}
\par
The mass and width are plotted for $0<\theta_0<\pi$ in Fig.~\ref{m0g0}, from which
it appears that the Breit-Wigner approximation $\Gamma\ll m$ holds for $\theta_0$ close to $\pi$.
The specific case considered in Refs.~\cite{Calmet,Casadio} is characterised by
$m\simeq\Gamma$ and precisely occurs at $m\approx\mpl/2$. 
The mass and width are also plotted for the next sheet, with $2\,\pi<\theta_1<3\,\pi$ in Fig.~\ref{m1g1},
and a similar behaviour appears in higher sheets: the maximum value of the mass $m$ and
width $\Gamma$ decrease with $n$, and the Breit-Wigner approximation holds for (relatively) large
$\theta_n$.
\subsection{Riemann sheets}
It is now interesting to look at the mass $m$ and width $\Gamma$ as functions of the renormalisation
scale $\mu$.
This can be done by expressing $\mu$ as a function of the phase $\theta_n$ from Eq.~(\ref{theta})
and then plotting $m$ and $\Gamma$ parametrically with respect to $\mu$.
For the fundamental sheet with $n=0$, the mass $m$ and the width $\Gamma$ are shown in Fig~\ref{mGamVmu0}
and their ratio in Fig.~\ref{RatioVmu0}.
From these graphs, we see again that $m\simeq\Gamma\simeq \mu\simeq\mpl$, in agreement
with Refs.~\cite{Calmet,Casadio}.
\par
\begin{figure}[t]
\centerline{\epsfxsize 8cm \epsfbox{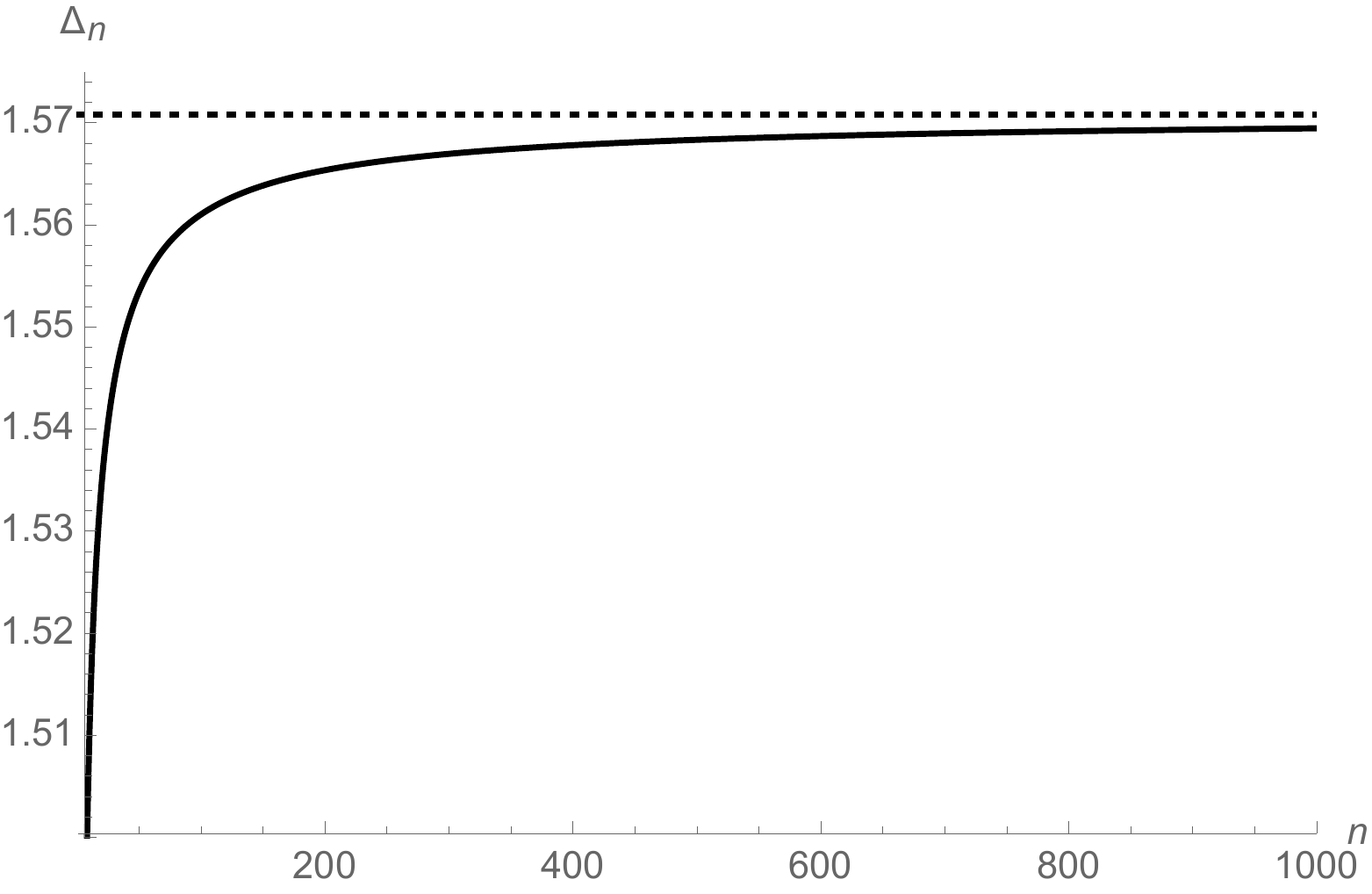}}
\caption{Phase $\Delta_n\equiv \theta_n-2\,\pi\,n$ for $\mu=m_{\rm P}$ and $n=1$ to $n=1000$ (solid line).
The dotted line is the asymptote $\pi/2$. 
\label{shift1}}
\end{figure}
\begin{figure}[t]
\centerline{\epsfxsize 8cm \epsfbox{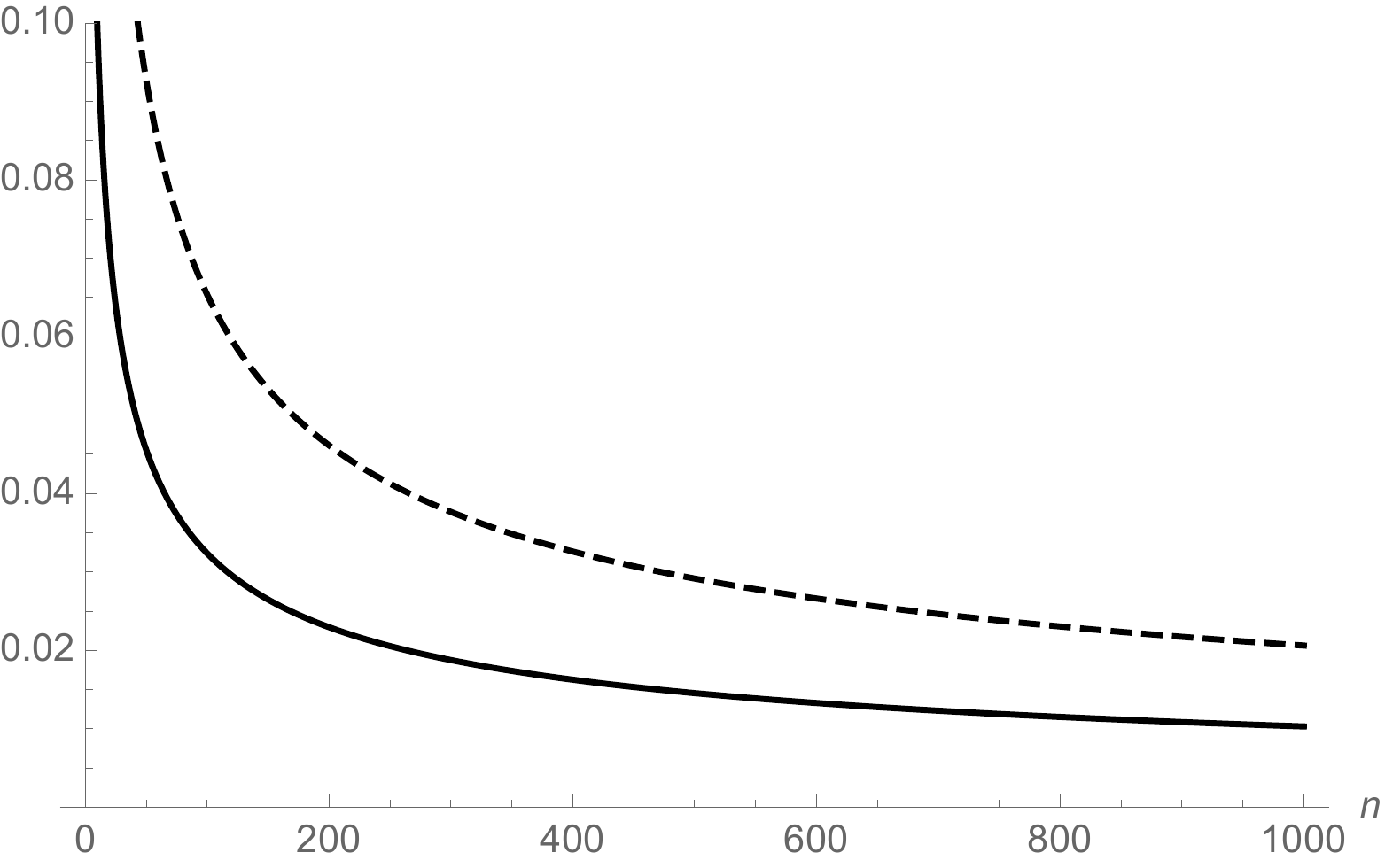}}
\caption{Mass $m$ (solid line) and width $\Gamma$ (dashed line) for $\mu=m_{\rm P}$ and
$n=1$ to $n=1000$.
\label{mGmu1}}
\end{figure}
\begin{figure}[t]
\centerline{\epsfxsize 8cm \epsfbox{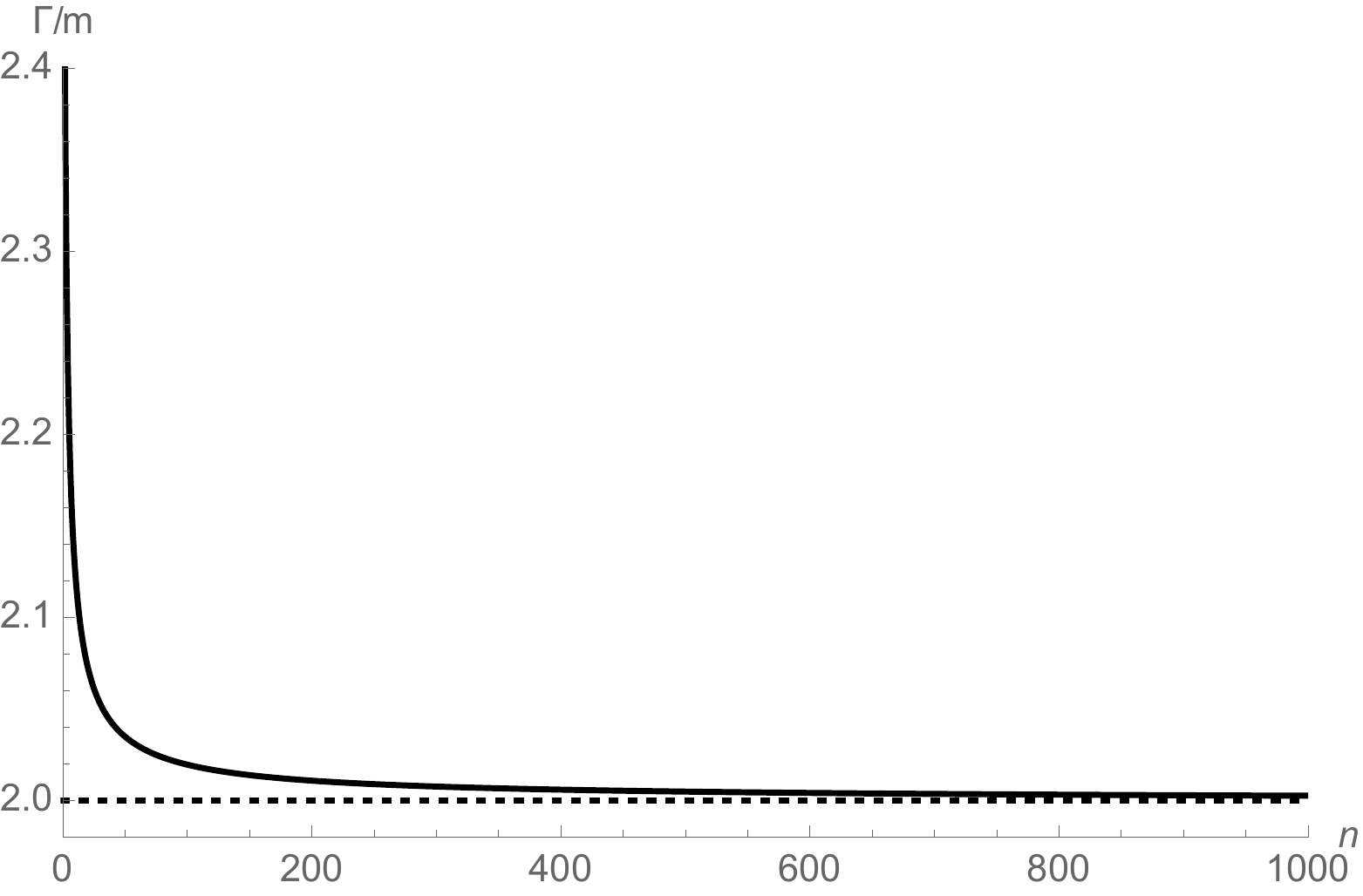}}
\caption{Ratio $\Gamma/m$ for $\mu=m_{\rm P}$ and $n=1$ to $n=1000$. 
The asymptote is $\Gamma/m=2$ (dotted line).
\label{ratiomu1}}
\end{figure}
\begin{figure}[t]
\centerline{\epsfxsize 8cm \epsfbox{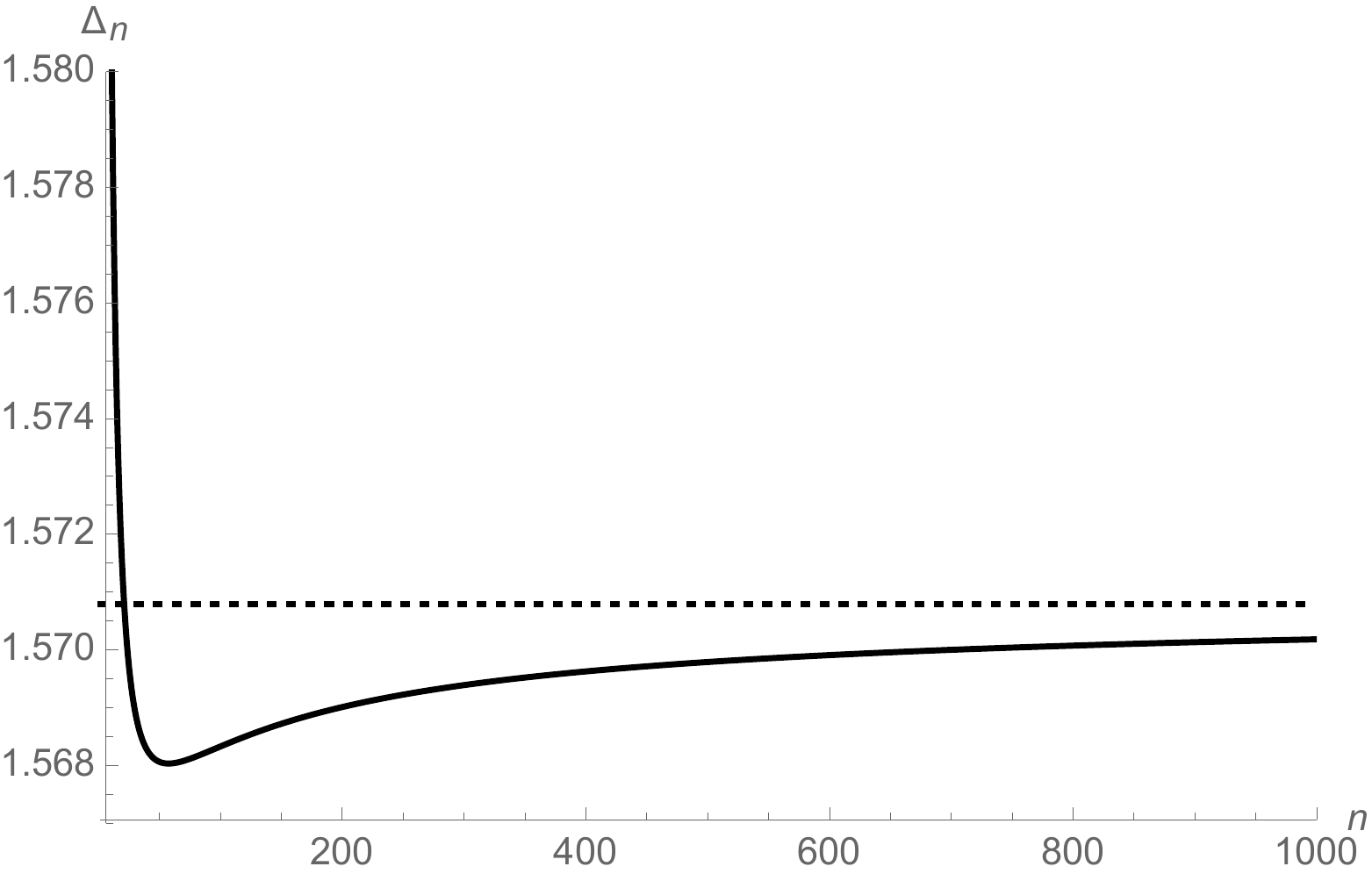}}
\caption{Phase $\Delta_n\equiv \theta_n-2\,\pi\,n$ for $\mu=m_{\rm P}/10$ and $n=1$ to $n=1000$ (solid line).
The dotted line is the asymptote $\pi/2$. 
\label{shift1s10}}
\end{figure}
\begin{figure}[t]
\centerline{\epsfxsize 8cm \epsfbox{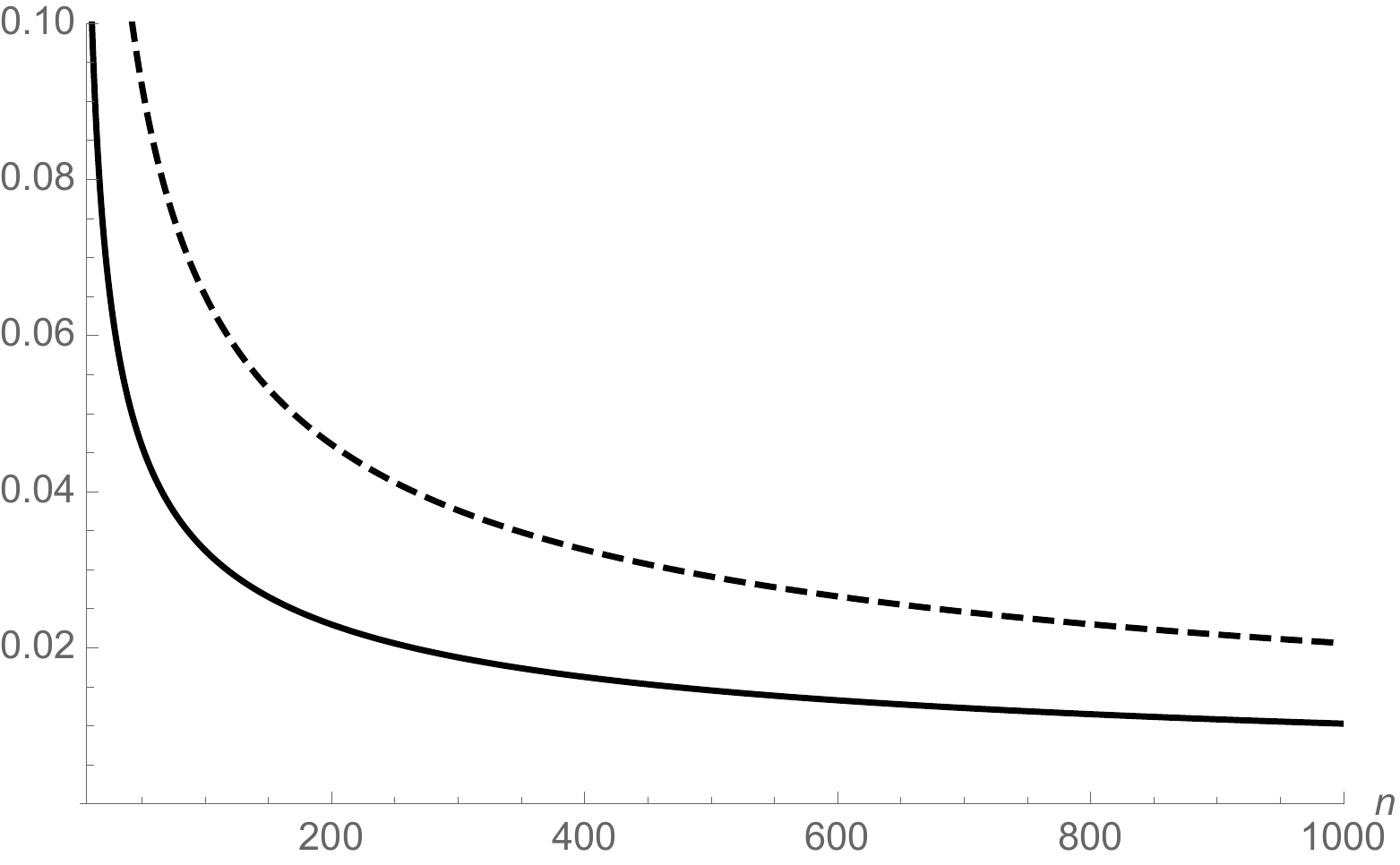}}
\caption{Mass $m$ (solid line) and width $\Gamma$ (dashed line) for $\mu=m_{\rm P}/10$ and
$n=1$ to $n=1000$.
\label{mGmu1s10}}
\end{figure}
\begin{figure}[t]
\centerline{\epsfxsize 8cm \epsfbox{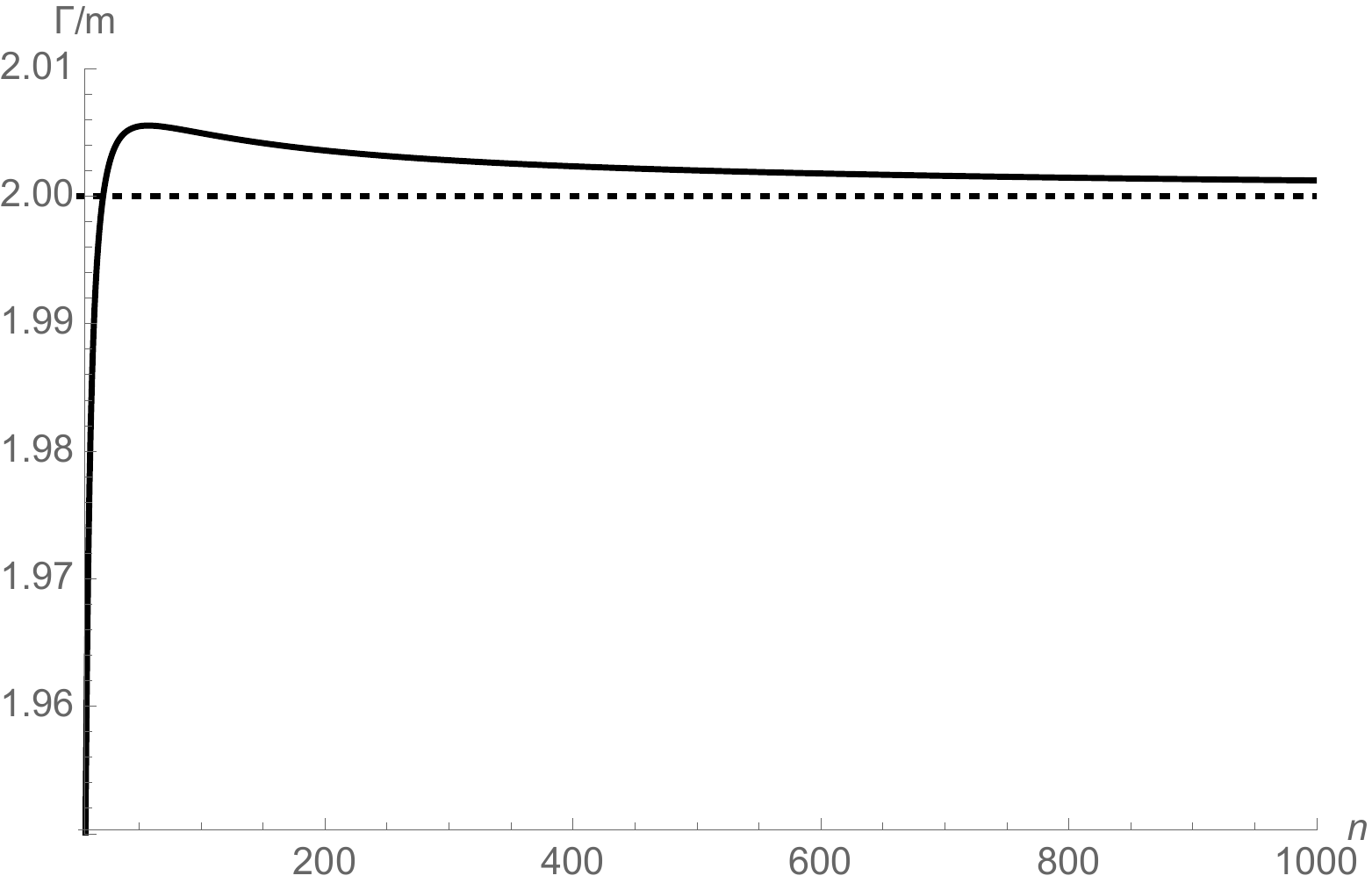}}
\caption{Ratio $\Gamma/m$ for $\mu=m_{\rm P}/10$ and $n=1$ to $n=1000$. 
The asymptote is $\Gamma/m=2$ (dotted line).
\label{ratiomu1s10}}
\end{figure}

For describing the situation in sheets with $n>0$, we find it convenient to fix the value of the renormalisation
scale $\mu$, hence $A$, and show the dependence of the phase $\theta_n$, mass $m$ and width $\Gamma$
on $n$.
For $\mu=m_{\rm P}$, the difference $\Delta_n\equiv \theta_n-2\,\pi\,n$ is plotted in Fig.~\ref{shift1}, from which
we see that $\Delta_n$ starts out smaller than $\pi/2$ and then asymptotes to $\pi/2$ monotonically from below.
The corresponding mass $m$ and width $\Gamma$ are shown to decrease monotonically in Fig.~\ref{mGmu1},
and their ratio (displayed in Fig.~\ref{ratiomu1}) asymptotes from above to $\Gamma/m=2$, in agreement with
Eq.~\eqref{R}.
For $\mu=m_{\rm P}/10$, the picture changes slightly.
The shift $\Delta_n$ is initially larger than $\pi/2$ but decreases below it for increasing $n$, and eventually
asymptotes to $\pi/2$ again from below (see Fig.~\ref{shift1s10}).
The mass and width in Fig.~\ref{mGmu1s10} behave qualitatively the same as before, but their
ratio starts out smaller than $\Gamma/m=2$, crosses over and then asymptotes to
$\Gamma/m=2$ again from above (see Fig.~\ref{ratiomu1s10}). 
\par
These behaviours can in fact be explained analytically from Eq.~\eqref{theta}.
Let us write $\Delta_n=\delta_n+\pi/2$, so that
\begin{equation}
\theta_n
=
2\,\pi\,n+\pi/2+\delta_n
\equiv \bar\theta_{n}+\delta_n
\ .
\label{dn}
\end{equation}
For $n\gg 1$, we can then see that $|\delta_n|\ll 1$. 
In fact, upon replacing~\eqref{dn} into Eq.~\eqref{theta}, we obtain
\begin{equation}
\bar\theta_n\,
e^{-\bar\theta_n\,\delta_n}
\simeq
A
\ ,
\end{equation}
to leading order in $\delta_n$, that is
\begin{equation}
\delta_n
\simeq
\frac{1}{\bar\theta_n}\,
\ln\!\left(\frac{A}{\bar\theta_n}\right)
\ .
\end{equation}
Since $\bar\theta_n$ will become necessarily larger than $A$, for sufficiently large $n$, the
logarithm will become negative and $\delta_n<0$ will asymptotically vanish. 
This clearly explains why $\Delta_n$ always asymptotes to $\pi/2$ from below.
Moreover the ratio $\Gamma/m$ asymptotes to 2 because of Eq.~\eqref{R} and does so from
above because $\delta_n<0$ for sufficiently large $n$.
Finally, let us note that for large $n$ we have
\begin{equation}
m
\sim
\Gamma
\sim
\frac{1}{\sqrt{\bar \theta_n}}
\sim
\frac{1}{\sqrt{n}}
\ .
\end{equation}
\section{Physical interpretation}
In this paper we have studied the analytic structure of the resummed graviton propagator~\cite{grav-prop,grav-prop1},
inspired by the possibie existence of black hole precursors~\cite{Calmet,Casadio}.
Remarkably, this structure depends essentially on the value of  the renormalisation scale parameter $\mu^2$.
It is instructive to compare this situation with the more studied cases of the
QED and the QCD.
In QED, the choice of the renormalisation point is conveniently performed in the classical limit, when the value of the
charge is extracted from macroscopic measurements.
At the same time it is possible to trade the dependence on the normalisation point $\mu^2$ in favour of the parameter
$\Lambda_{\rm QED}$, which is connected with the Landau pole~\cite{AKL}, whose value is very large.
In QCD, the analogous parameter $\Lambda_{\rm QCD}$ is typically of hadronic mass scale.
Since QCD does not admit a classical limit, this parameter is the one used for describing experimental
data~\cite{Shirkov1,anal-other}.   
\par
Since gravity is not renormalisable, it must be treated as an effective theory with no parameter equivalent to 
$\Lambda$, and the dependence on the renormalisation scale $\mu^2$ cannot be avoided.  
In order to have stable black hole-like quasi-particles in the spectrum, one would need widths $\Gamma\ll m$,
which our analyses in turn showed would require a renormalisation scale $\mu^2\ll m_{\rm P}^2$.
In any case, the number of such states would be finite, since the ratio $\Gamma/m$ grows with the sheet
number $n$ towards the asymptotic value $\Gamma/m=2$.
On the other hand, the existence of such quasi-particle states with positive $\Gamma$ implies the existence
of equal mass states with negative $\Gamma$.
One might speculate these states would be white hole precursors, although this physical interpretation is not
completely clear to us.
Perhaps highly unstable black and white hole-like states in higher Riemann sheets are not dangerous 
because of their highly virtual character and the cancellation of the corresponding imaginary parts
in the amplitudes involving the graviton propagator.   
One should mention here that complex conjugated singularities do appear in the Gribov theory of quark
confinement~\cite{Gribov}.
They appear in the solutions of the Schwinger-Dyson equation in QCD~\cite{Roberts, Maris}
and also in the extra-dimensional field theories~\cite{Kazakov:2007su}.
\par
It is perhaps more compelling to consider the lower Riemann sheets, where poles with positive mass
come likewise with both positive and negative width.
In order to deal with the states with negative $\Gamma$ in lower Riemann sheets, one can consider (at least)
two options.
The first one is to define the contour of integration for computing the propagator in position space in such
a way as to exclude these poles, in the spirit of Refs.~\cite{Redmond,Shirkov}.
In that case, the extra imaginary phase due to the black hole width might lead to observable effects,
at least in principle, in particular, to single spin asymmetries~\cite{TY}. 
Another option is to assume the renormalisation scale $\mu^2\gtrsim m_{\rm P}^2$, so that $|\Gamma|\gtrsim m$
and these states appear only as virtual particles. 
We conclude by saying that  the connection between the black hole type states and the effective quantum field theory
deserves further investigation.
For instance, non-perturbative effects might come into play at scales of the order of $m_{\rm P}$, which
would require a different analysis like the one in Ref.~\cite{Casadio}. 
\section*{Acknowledgments.}
We are indebted to D.~Kazakov, E.~Kolomeitsev, C.~Roberts, G.~Venturi and R.~Yahibbaev
for useful discussions.
O.T.~would like to thank INFN for kind hospitality during October 2016. 
X.C.~is supported in part by the Science and Technology Facilities Council
grant N. ST/J000477/1. 
R.C.~and A.K.~are partly supported by the INFN grant FLAG.
A.K.~was partially supported by the RFBR grant N.~17-02-01008. 
O.T.~was partially supported by the RFBR grant N.~17-02-01108.
\appendix
\section{Gravitational action and propagator}
\label{A1}
As shown in \cite{XC}, the resummed propagator~\eqref{prop} can be seen as coming
from the variation with respect of the metric field $g_{\mu\nu}$ of the linearized version
of the effective action
\begin{eqnarray}\label{action1}
S
&\!\!=\!\!&
\int d^4x \,
\sqrt{-g}
\left[ \frac{\bar m_{\rm P}^2}{2}\,\mathcal{R}
+ c_1\, \mathcal{R}^2
+ c_2\, \mathcal{R}_{\mu\nu}\,\mathcal{R}^{\mu\nu}
\right.
\nonumber
\\
&&
\left.
+ b_1\, \mathcal{R}\, \log\!\left( \frac{\Box}{\mu^2}\right)\mathcal{R}
+ b_2\, \mathcal{R}_{\mu\nu}\,  \log\!\left( \frac{\Box}{\mu^2}\right)\mathcal{R}^{\mu\nu}  
+ b_3\,\mathcal{R}_{\mu\nu\rho\sigma}\,  \log\!\left( \frac{\Box}{\mu^2}\right)\mathcal{R}^{\mu\nu\rho\sigma}
\right]
\ , 
\end{eqnarray}
where $\mathcal{R}$, $\mathcal{R}^{\mu\nu}$ and $\mathcal{R}^{\mu\nu\rho\sigma}$
are respectively the Ricci scalar, Ricci tensor and Riemann tensor and $\bar m_{\rm P}$
is the reduced Planck mass.
The Wilson coefficients $c_1$ and $c_2$ are arbitrary within the effective field theory approach,
and should be fixed by comparing with experimental data.
On the other hand $b_1$, $b_2$ and $b_3$ are calculable from first principles and related to $N$.
\par
The resulting complete propagator for the graviton then contains the function
\begin{equation}
G^{-1}(p^2)
=
2\,p^2
\left[1-  16\, \pi\, c_2 \,\frac{p^2}{\mpl^2}
-\frac{N\,p^2}{120\,\pi\, \mpl^2}\ln\!\left(-\frac{p^2}{\mu^2}\right)\right]
\ .
\label{prop3}
\end{equation}
Clearly the position of the poles of~\eqref{prop3} will depend on the value of $c_2$,
which is arbitrary.
For the self-healing mechanism to work, $c_2=c_2(\mu)$ should be suppressed by $1/N$
in comparison to the coefficients $b_i$~\cite{Tomboulis:1977jk}.
Let us also stress that Eq.~\eqref{prop3} can be formally rewritten in the same form as
Eq.~\eqref{prop}, namely
\begin{equation}
G^{-1}(p^2)
=
2\,p^2
\left[1-\frac{N\,p^2}{120\,\pi\, \mpl^2}\ln\!\left(-e^{1920\,\pi^2\,c_2}\,\frac{p^2}{\mu^2}\right)\right]
\ ,
\label{prop2}
\end{equation}
so that the analytic structure does not change and our analysis still applies.

\begin{thebibliography}{99}
%
%
\bibitem{Holstein}
B.R.~Holstein,
{\em Topics in Advanced Quantum Mechanics},
(Dover, New York, 2014).
%
\bibitem{Bog-Shir}
N.N.~Bogoliubov, D.V.~Shirkov,
{\em Introduction to the theory of quantized fields}, 
(John Wiley, New York, 1980).
%
\bibitem{Ber-Lif-Pit}
V.B.~Berestetskii, E.M.~Lifshitz and L.P.~Pitaevskii, 
{\em Quantum Electrodynamics: Volume 4},
(Elsevier Science, Amsterdam, 1982).
%
\bibitem{AKL}
L.D.~Landau, A.A.~Abrikosov and I.M.~Khalatnikov,
Dokl. Akad. Nauk SSSR 95 (1954) 773;
Dokl. Akad. Nauk SSSR 95 (1954) 1177;
Dokl. Akad. Nauk SSSR 96 (1954) 261. 
%
\bibitem{Redmond}
P.J.~Redmond,
Phys.\ Rev.\  {\bf 112} (1958) 1404.
%
\bibitem{Shirkov}
N.N.~Bogolyubov, A.A.~Logunov and D.V.~Shirkov,
Sov. Phys. JETP 37 (1960) 574.
%
\bibitem{Shirkov1}
D.V.~Shirkov and I.L.~Solovtsov,
Phys.\ Rev.\ Lett.\   79 (1997) 1209;
Theor.\ Math.\ Phys.\  {\bf 150} (2007) 132.
%
\bibitem{anal-other}
A.P.~Bakulev, S.V.~Mikhailov and N.G.~Stefanis,
Phys.\ Rev.\ D {\bf 72} (2005) 074014;
M.~Baldicchi, A.V.~Nesterenko, G.M.~Prosperi, D.V.~Shirkov and C.~Simolo,
Phys.\ Rev.\ Lett.\  {\bf 99} (2007) 242001;
R.S.~Pasechnik, D.V.~Shirkov and O.V.~Teryaev,
Phys.\ Rev.\ D {\bf 78} (2008) 071902
A.P.~Bakulev, S.V.~Mikhailov and N.G.~Stefanis,
JHEP {\bf 1006} (2010) 085;
V.L.~Khandramai, R.S.~Pasechnik, D.V.~Shirkov, O.P.~Solovtsova and O.V.~Teryaev,
Phys.\ Lett.\ B {\bf 706} (2012) 340.
%
\bibitem{grav-prop}
T.~Han and S.~Willenbrock,
Phys.\ Lett.\ B 616 (2005) 215.
%
\bibitem{grav-prop1}
U.~Aydemir, M.M.~Anber and J.F.~Donoghue,
Phys.\ Rev.\ D 86 (2012) 014025.
%
\bibitem{Calmet}
X.~Calmet,
Mod.\ Phys.\ Lett.\ A {\bf 29} (2014) 1450204.
%
\bibitem{Casadio}
X.~Calmet and R.~Casadio,
Eur.\ Phys.\ J.\ C {\bf 75} (2015) 445.
%
\bibitem{Lambert}
R.M.~Corless, G.H.~Gonnet, D.E.G.~Hare, D.J.~Jeffrey and D.E.~Knuth,
Adv.\ Comput.\ Math.\  {\bf 5} (1996) 329.
%
\bibitem{BW}
G.~Breit and E.~Wigner,
Phys.\ Rev.\  {\bf 49} (1936) 519.
%
\bibitem{wille}
T.~Bhattacharya and S.~Willenbrock,
Phys.\ Rev.\ D {\bf 47} (1993) 4022.
%
%
\bibitem{bohm}
A.R.~Bohm and Y.~Sato,
Phys.\ Rev.\ D {\bf 71}, 085018 (2005).
%
\bibitem{Gribov}
V.N.~Gribov,
in Nyiri, J. (ed.), {\em The Gribov theory of quark confinement},
162-184 [hep-ph/9403218];
in Nyiri, J. (ed.), {\em The Gribov theory of quark confinement},
185-191 [hep-ph/9404332];
in Nyiri, J. (ed.), {\em The Gribov theory of quark confinement},
192-203 [hep-ph/9905285];
Eur.\ Phys.\ J.\ C {\bf 10} (1999) 91.
%
\bibitem{Roberts}
G.~Krein, C.D.~Roberts and A.G.~Williams,
Int.\ J.\ Mod.\ Phys.\ A {\bf 7} (1992) 5607;
C.J.~Burden, C.D.~Roberts and A.G.~Williams,
Phys.\ Lett.\ B {\bf 285} (1992) 347.
%
\bibitem{Maris}
P.~Maris,
Phys.\ Rev.\ D {\bf 52} (1995) 6087
%
\bibitem{Kazakov:2007su}
D.I.~Kazakov and G.S.~Vartanov,
JHEP {\bf 0706} (2007) 081.
  %
\bibitem{TY}
O.V.~Teryaev, R.M.~Yahibbaev,
work in progress. 
%
\bibitem{XC}
X.~Calmet, S.~Capozziello and D.~Pryer, in preparation. 
%
\bibitem{Tomboulis:1977jk}
E.~Tomboulis,
Phys.\ Lett.\  {\bf 70B} (1977) 361.
%
%
\end{thebibliography}
\end{document}